\documentclass[prl,showpacs,floatfix,twocolumn]{revtex4}

\usepackage{bm}
\usepackage{graphics}
\usepackage{graphicx}
\usepackage[dvips,usenames]{color}

\newcommand{\av}[1]{{\langle{#1}\rangle}}

\newcommand{\be}{\begin{equation}}
\newcommand{\ee}[1]{\label{#1} \end{equation}}
\newcommand{\ba}{\begin{eqnarray}}
\newcommand{\ea}[1]{\label{#1} \end{eqnarray}}
\newcommand{\nl}{\nonumber \\}

\newcommand{\AuthorTeam}{
\author{T.~S.~Bir\'o\footnote{tsbiro@sunserv.kfki.hu}} 
\affiliation{
 KFKI Research Institute for Particle and Nuclear Physics,
 H-1525 Budapest, P.O.Box 49, Hungary
}
\author{P.~L\'evai} 
\affiliation{
 KFKI Research Institute for Particle and Nuclear Physics,
 H-1525 Budapest, P.O.Box 49, Hungary
}
\author{P.~V\'an}
\affiliation{
 KFKI Research Institute for Particle and Nuclear Physics,
 H-1525 Budapest, P.O.Box 49, Hungary
}
\author{J.~Zim\'anyi}
\affiliation{
 KFKI Research Institute for Particle and Nuclear Physics,
 H-1525 Budapest, P.O.Box 49, Hungary
}
}

\begin{document}

\title{{ Mass distribution from quark matter equation of state}}

\AuthorTeam

\date{{\bf \today}}

\begin{abstract}
We analyze  the equation of state in terms of 
quasiparticles with continuously distributed mass.
We seek for a description of the entire pressure -- temperature curve  
at vanishing chemical potential 
in terms of a temperature independent mass distribution. 
We point out properties indicating a mass gap in this 
distribution, conjectured to be related to confinement.
\end{abstract}

\maketitle





According to the proposal of A. Jaffe and E. Witten,  
a successful quantum Yang-Mills theory must have a mass gap~\cite{JaffeWitten}.
In heavy ion collisions 
a deconfined phase is expected to form, and the produced
quark-gluon plasma (QGP) is described by Quantum Chromodynamics (QCD).
Thus one can expect the appearance of such a mass-gap in the spectral function
of the basic QGP degrees of freedom, namely quarks and gluons.
In this paper we perform
a quantitative analysis on some results from lattice QCD on the equation 
of state (eos) and reconstruct it from a mass distribution
of non-interacting quasi-particles.
We present strong indications for a mass gap in this distribution. 

We have used earlier a mass distribution for massive quarks
and developed a coalescence picture~\cite{JPG} to describe hadronization 
of deconfined quark matter, and reproduced final state hadron
ratios and transverse spectra successfully. That model was  based 
on an earlier coalescence model~\cite{ALCOR,COALESCE}, where quarks and gluons had 
finite effective masses without any width.
The apparent entropy reduction problem by coalescence with an associated reduction
(confinement) of color degrees of freedom can be resolved by assuming sufficiently
massive partons around the hadronization temperature in the precursor matter.
The necessary mass scale for quarks is about $300-350$ MeV and even higher (about $700$ MeV)
for gluons~\cite{MASSIVE-GLUON}, thus we could assume 
that in the prehadronization stage the
heavy gluons decay into quark -- antiquark pairs~\cite{ALCOR}.
Recently, partonic level models of heavy ion reactions
also utilized the quark coalescence picture successfully~\cite{COA2,COA3}.

Considering quark coalescence as the mechanism of hadronization one has to deal with the question,
how to make a hadron with a  mass  lower than the sum of two parton masses.
In order to solve this problem we have introduced distributed mass partons into our
hadronization model~\cite{JPG}.
Having in medium partons in quark matter as precursors of emerging
hadrons in mind, we connect now the distributed mass
parton picture to a simplified treatment of spectral functions.

The equation of state of an interacting system, when analyzed in terms of quasiparticles,
is coded in the spectral function $\rho(\omega,\vec{p})$. 
Our ansatz to this assumes a particular form: 
\be
 \rho(\omega,\vec{p}) = 2\pi \frac{w(m)}{2m} \Theta(m^2) \left( \Theta(\omega)-\Theta(-\omega) \right)
\ee{OUR-SP-ANSATZ}
with $m^2=\omega^2-\vec{p}^2$ and $\Theta(x)$ being the step function.
A continuous $w(m)$ mass distribution describes a finite width ansatz for the spectral function.
The normalization of the spectral function,
$\int_{-\infty}^{+\infty} \rho(\omega,\vec{p}) \omega d\omega/\pi = 1$, requires 
$\int_{0}^{\infty} w(m) dm = 1$.
The conventional quasiparticle approach on the other hand often explores a Breit-Wigner form of the
spectral density \cite{PeshierSP},
\be
 \rho(\omega,\vec{p}) = \frac{\gamma}{E} 
 \left( \frac{1}{(\omega-E)^2+\gamma^2} - \frac{1}{(\omega+E)^2+\gamma^2} \right)
\ee{PESHIER-SP}
with $E^2=M^2-\gamma^2+\vec{p}^2$ and temperature dependent parameters $M(T)$
and $\gamma(T)$. 



Thermodynamical consistency of the quasiparticle picture imposes 
constraints on the mass distribution, $w(m)$,  
in particular on its dependence on the temperature
or on other medium parameters~\cite{TONEEV}. In this letter we investigate the
possibility of
a temperature independent mass distribution and therefore neglect the mean field
term for consistency.
The total pressure at vanishing chemical potential is given as the following
integral: 
\be
  p(T) = \int_0^{\infty} w(m) \, p(m,T) dm.
\ee{TOTAL-PRESS}
One may suppose that only a single mass scale occurs in the mass distribution,
so it can be expressed by a dimensionless distribution:
\be
 w(m) = \frac{1}{T_c} f(\frac{m}{T_c}).
\ee{PARTICULAR-MASS-DISTRIBUTION}
The normalization integral for $w$ is inherited by the shape (form factor)
function $f(t)$:
\be
 \int_0^{\infty} w(m) dm = \int_0^{\infty} f(t) dt = 1.
\ee{SPEC-MASS-INTEGRAL}

The quark gluon plasma at vanishing chemical potential has the pressure
\be
 p(T) \: = \: \sigma(z) \kappa T^4,
\ee{SPEC-PRESSURE}
with $z=T_c/T$ and $\kappa$ being the Stefan-Boltzmann constant. 
In the Boltzmann approximation 
the fixed $m$-contributions are given by the Bessel K-function,
$p(m,T) \propto T^4 \Phi(m/T)$ 
with\footnote{The coefficients have been chosen here so that $\Phi(0)=1$.}
$\Phi(u)=u^2 K_2(u)/2$.
Deviations in 
$\Phi(m/T) = p(m,T)/p(0,T)$ due to using Bose or Fermi distributions as a function
of $m/T$ never exceed six per cent.
Thus in the distributed mass model 
the $\sigma(z)$ function in this approximation is given by the integral
\be
 \sigma(z) \: = \: \int_0^{\infty} f(t) \: \frac{(zt)^2}{2} K_2(zt) \, dt.
\ee{SIGMA}
This may be recognized as the so called Meijer \mbox{K-transform}~\cite{ERDELYI,MEIJER} 
(a generalized Laplace transform)
of the $ f(t) $ function. The inverse of 
this transformation yields the mass distribution function in terms of the
observed $ \sigma(z) $ values : 
\be
 f(t) \: = \: \frac{2}{i\pi } \int_{c-i\infty}^{c+i\infty} \sigma(z) \: \frac{I_2(zt)}{zt} \, dz.
\ee{INV-MEIJER}
This raises a peculiar question: is it possible to show, that to any $\sigma(z)$ function
extracted from an equation of state (e.g. from lattice QCD calculations) there exists a unique
mass distribution $f(t)$ with the mass scale parameter kept
temperature and chemical potential independent? In this case the shape of the mass distribution
is not arbitrary. 
Of course, the Meijer K-transform is invertible, but
one has to check whether the $ f(t) $ function obtained by eq.(\ref{INV-MEIJER}) is positive
semidefinit and normalized to unity. The normalization is the easier problem,
the $\sigma(0)$ limit being directly the integral of the  $ f(t) $ function due to
the small argument behavior of the Bessel K-function. 
It can, however, be difficult to arrive at a nowhere negative $ f(t) $ by
knowing  $\sigma(z)$ only at some points on the real $z$-axis. 

Before investigating any particular ansatz for $\sigma(z)$ let us consider an important
general property. There is a relation between the integration moments of this
quantity (the scaled pressure) and the normalized mass distribution, $f(t)$:
\be
 M_n = \int_0^{\infty} \! z^{n-1} \sigma(z) dz \: = \: I_n
 \int_0^{\infty} f(t) t^{-n}  dt
\ee{MASS-MOMENTS}
with
\be
 I_n = \frac{1}{2} \int_0^{\infty} u^{n+1} K_2(u) du \: = \: 
 2^n \Gamma\left(2+\frac{n}{2} \right) \Gamma\left(\frac{n}{2} \right)
\ee{COEFF}
where $\Gamma(x)$ is Euler's Gamma function. This is finite for positive $n$
and divergent for zero or negative integer values. We conclude that as long as the
$M_n$ moments of the eos curve are finite so must be the inverse mass
moments of the mass distribution. Since due to construction $\sigma(0)=1$ and
$\sigma(z)$ is rapidly decreasing due to confinement for large $z=T_c/T$ (low temperature),
any mass distribution reconstructing the equation of state of QCD must be suppressed
for low masses. 

\begin{figure}
\begin{center}
 \includegraphics[width=0.33\textwidth,angle=-90]{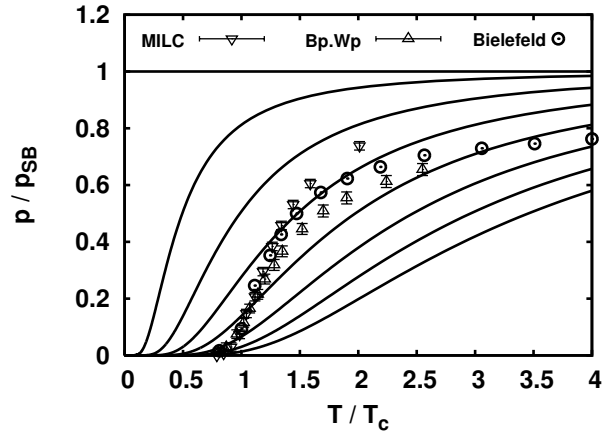}
 \end{center}

 \caption{ \label{FIG-FIXMASS}  
  The pressure normalized to the massless Stefan-Boltzmann value as a
  function of the scaled temperature $T/T_c$ for different constant mass relativistic gases (full lines
  in order for $M/T_c=0, 1, 2, 3, 4, 5, 6$ and $7$) 
  and from lattice QCD data of Ref.\cite{FODOR} (down triangles),
  of Ref.\cite{BIELEFELD} (open circles) and of Ref.\cite{MILC} (up triangles). 
 }
 \end{figure}

Let us now discuss how to obtain a particular functional form for $\sigma(z)$.
The high-temperature (small $z$) expansion of $\Phi(zt)$ leads to
\ba
 \frac{p(T)}{\kappa T^4} \: &=& \: 1 - \frac{\av{m^2}}{4T^2} 
  +\frac{\av{m^4}}{16T^4}\left(\frac{3}{4}-\gamma\right) +\frac{\av{m^4\ln\frac{2T}{m}}}{16T^4}    \nl
 & &+ \frac{\av{m^6}}{192T^6}\left(\frac{17}{12}-\gamma\right) + \frac{\av{m^6\ln\frac{2T}{m}}}{192T^6} \: + \: \ldots
\ea{highT-sigma}
with $\gamma$ being the Euler-Mascheroni  constant.
Accidentally the perturbative QCD pressure (taken as the finite part at the scale $2\pi T$)
shows a similar structure,
\be
 \frac{p(T)}{\kappa T^4} \: = \: 1 - a_2g^2 + a_4g^4 + b_4g^4\ln \frac{2\pi T}{\Lambda} + \ldots
\ee{pQCD-naive}
($a_2\approx 0.072, a_4\approx 0.061, b_4\approx 0.008$ for $N_f=3$ based on Ref.\cite{Kajantie}).
It is possible to fit this form as a high-T asymptotics by assuming a scaling of
expectation values, like $\av{m^2}=cg^2T^2$, $\av{m^4}=c'g^4T^4$, etc.
This is the basis of the traditional quasiparticle picture\cite{PESHIER}, at the same time
it predicts a width changing with the temperature. This view assumes a temperature-dependent
mass distribution, $w(m,T)$, which - for the sake of thermodynamical consistency - would require
a temperature-dependent mean field pressure, $-B(T)$ to be taken into account.

There is, however, another possibility, which we would like to pursue in the present article.
The low-argument expansion eq.(\ref{highT-sigma}) fails if the expectation values, like
$\av{m^2}$, $\av{m^4}$, etc. are divergent. In fact this assumes a high-mass tail of the
$w(m)$ distribution not decaying faster than $~m^{-3}$. As we shall point out later, our numerical
efforts to obtain $w(m)$ agree with this statement.

Fig.\ref{FIG-FIXMASS} presents the normalized pressure for relativistic Boltzmann gases
with several fixed masses. The lattice QCD eos data of the Budapest-Wuppertal group~\cite{FODOR}
and of the Bielefeld group~\cite{BIELEFELD} seem to lie everywhere below the curve for
the mass $M=3T_c$, recent MILC data~\cite{MILC} below the curve for $M=2.5T_c$.
As a consequence, if one accepts this property also for lower
temperatures where actually no reliable simulations are available, the mass spectrum
$w(m)$ would not contain any mass lower than $M=3T_c$ or $M=2.5T_c$, respectively. 
This property can be verified by 
rigorous mathematical estimates of upper bounds for $w(m)$ on the interval 
$0<m<M$ \cite{MARKOV}.

The pressure of hot QCD has been recently calculated up to ${\cal O}(g^6\ln(1/g) )$
\cite{Kajantie}.
The result contains  formally $\ln(2\pi T/\Lambda)$ terms, but according to the
suggestion of the authors the coupling $g(\Lambda)$ should be taken at
$\Lambda\approx 6.47 T$ \cite{Kaj2} and this way the temperature dependence
of the normalized pressure stems from the temperature dependence of the coupling
strength, renormalized relative to a scale proportional to the temperature.
Agreement with lattice QCD eos data is achieved at the highest computable level
only, with an extra fit of a constant which is not calculable perturbatively.

\begin{figure}
\begin{center}
 \includegraphics[width=0.33\textwidth,angle=-90]{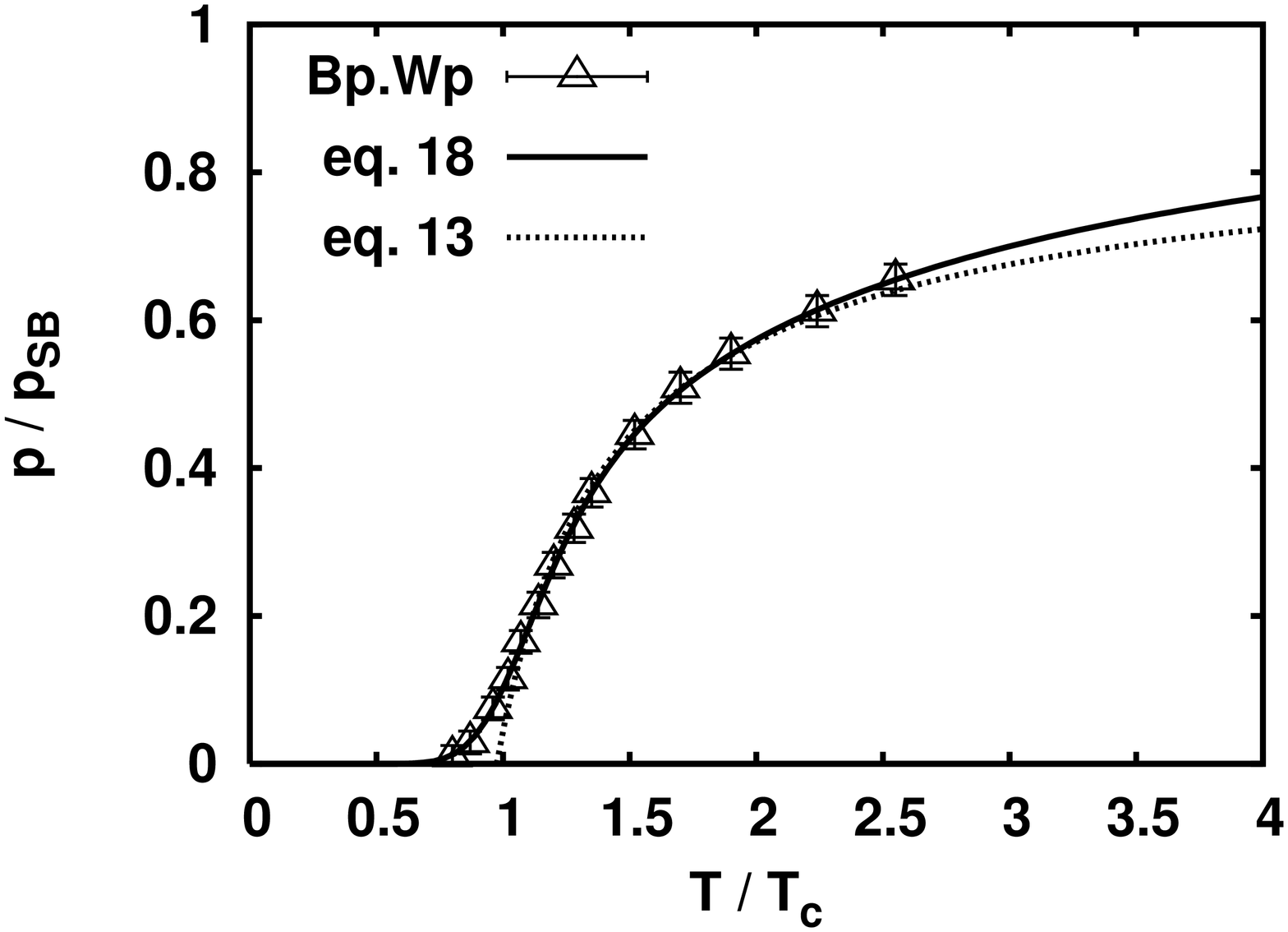}
 \includegraphics[width=0.33\textwidth,angle=-90]{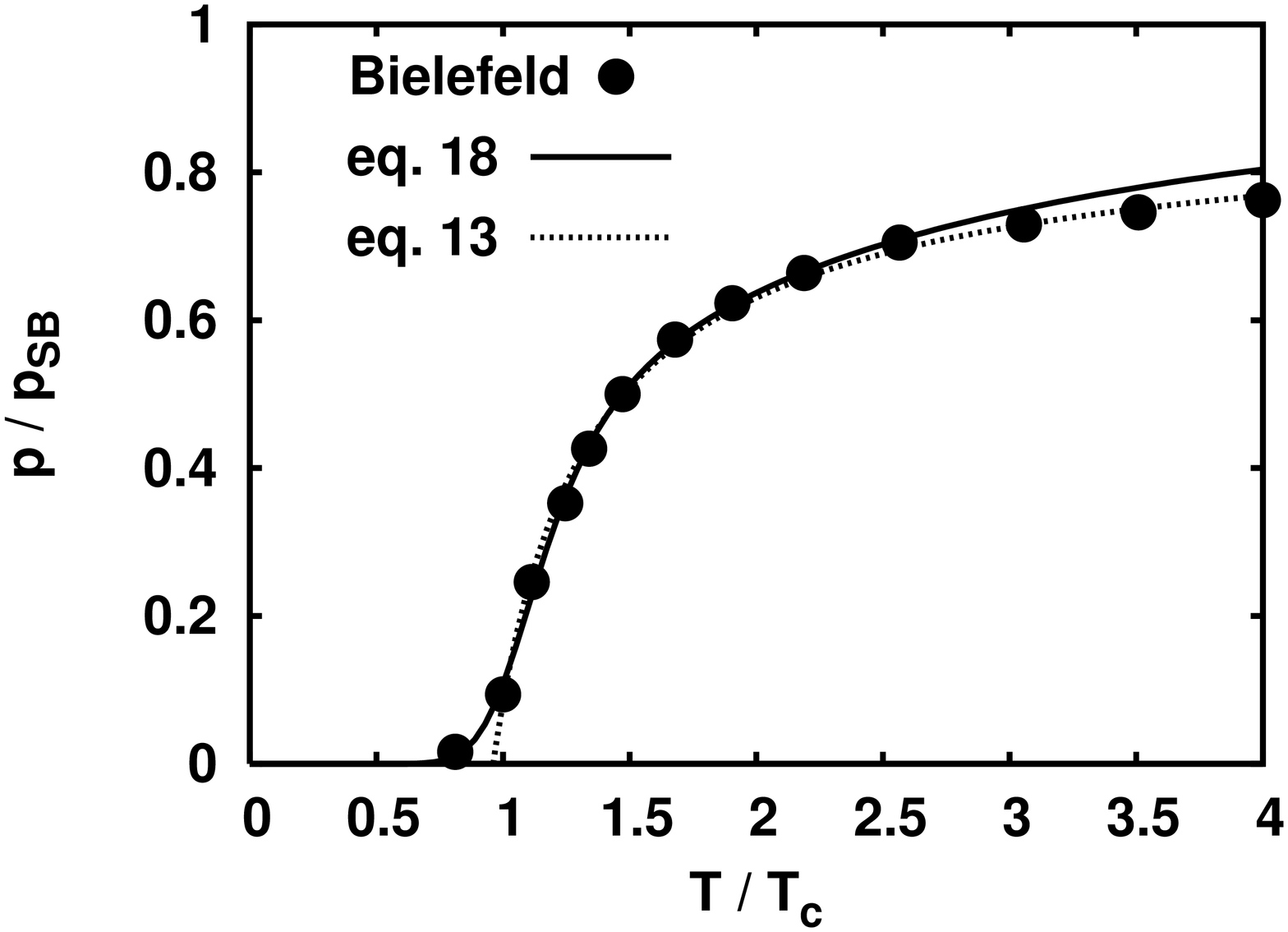}
 \includegraphics[width=0.33\textwidth,angle=-90]{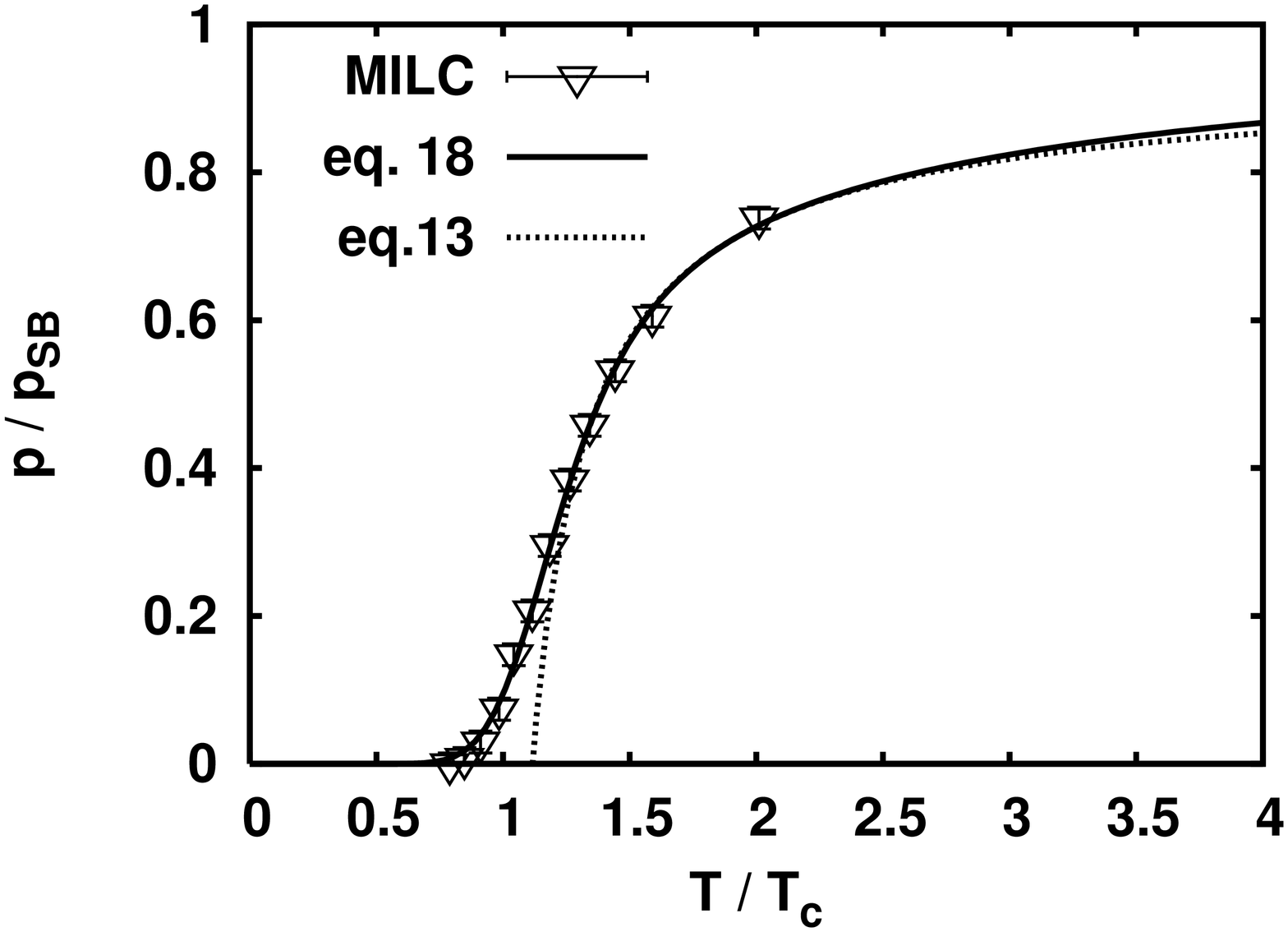}
 \end{center}

 \caption{ \label{FIG-EOS}  
  The lattice QCD pressure normalized to the massless Stefan-Boltzmann value as a
  function of the temperature $T$ from lattice eos results of Ref.\cite{FODOR}
 (above), Ref.\cite{BIELEFELD} (middle) and Ref.\cite{MILC} (bottom). 
  Our fits are indicated by the continuous lines, the $1-K/ln(\eta T/T_c)$-type
	fits by the dotted lines.
 }
 \end{figure}

It is intriguing that for practical purposes the ${\cal O}(g^2)$ formula by using
the 1-loop renormalized $g(T)=1/b\ln(T/\overline{\Lambda})$ can be also fitted to
lattice data by fitting $\overline{\Lambda}$. This results in a formula
\be
 \frac{p(T)}{\kappa T^4} \: = \: 1 - \frac{K}{\ln(\eta T/T_c)}
\ee{fitpQCD}
reaching zero pressure at $T \approx T_c$.
Fits to different lattice QCD equations of state leads to quantitative, but no
qualitative differences. For the data of Ref.\cite{FODOR} we obtain $K=0.54, \eta=1.76$,
for Ref.\cite{BIELEFELD} $K=0.43, \eta=1.6$ and for Ref.\cite{MILC} $K=0.22, \eta=1.12$
(cf. dotted lines on Fig.\ref{FIG-EOS}).

There are some theoretical signs on the other hand, that the $g(T)=1/b\ln(T/\Lambda)$ formula
is not necessarily the really high temperature limit prediction of QCD.
The finite temperature renormalization group result is of type\cite{RUNNING-ALFA}
\be
 \frac{1}{\alpha(Q^2,T^2)} \: = \: b \ln \frac{Q^2}{Q_0^2} + c \left(\frac{T^2}{Q^2}-\frac{T^2}{Q_0^2} \right)
\ee{finiteT-RG}
with $b=1/\alpha(Q_0^2,T^2)+b_0$ and $b_0=(11N_c/2-2N_f/3)/(4\pi)$ being the one-loop perturbative
beta function coefficient. This is usually considered in the $Q_0^2 \gg T^2$ limit and then,
assuming a sharp thermal distribution of $Q^2$ values, $Q^2=(a\pi T)^2$ is taken. 
This leads to the conjecture
\be
 \frac{1}{\alpha((a\pi T)^2,T^2)} = \frac{1}{\alpha(T^2)} = b_0 \ln \frac{T^2}{\Lambda^2}.
\ee{WRONG-ALFA}
Since $Q_0$ was large and $\Lambda$ is around $T_c$, also the coefficient $a$ is taken
as a large number. None of these assumptions is established by the QCD itself.
The assumption $Q_0^2 \gg T^2$  contradicts the $T \rightarrow \infty$ limit, the spread of a
thermal distribution of possible $Q^2$ values also increases like $T^2$, and finally
there is always a non-negligible influence of low $Q^2$ physics 
on the coupling at any finite temperature. 
In fact calculating the thermal distribution of $Q^2/T^2$ between two massless,
Boltzmann-distributed particles one obtains easily
\be
 P\left(\frac{Q^2}{T^2}\right) \: = \: \frac{1}{128}
 \left(\frac{Q^3}{T^3} K_1(\frac{Q}{T}) + 2 \frac{Q^2}{T^2} K_2(\frac{Q}{T}) \right).
\ee{Q2distr}
for $Q^2>0$. The probability of having $Q^2=0$ is finite at any temperature,
$P(0)=3/64$ and this is the maximum of $P(x)$.
As a consequence higher twist effects
which are not infrared safe (like $\ln Q^2$, $1/Q^2$ etc.) might destroy the
often quoted logarithmic scaling of the coupling constant with temperature.
It is probably the best to consider a $K/ln(\eta T/T_c)$-type formula (eq.\ref{fitpQCD})
as a standard, but not
the only possible fit to the lattice eos, with some parameters of nonperturbative origin.

In order to evaluate the inverse Meijer transform, eq.(\ref{INV-MEIJER}),
one has to approximate the lattice QCD data by an analytic $ \sigma(z) $ 
function. A family of mass distributions can be Meijer-transformed analytically:
\ba
w(m) &=& 
    \frac{1}{\Gamma(\nu)\Gamma(2-\nu)} \frac{2\lambda}{m^3} 
    \left(m^2-\lambda^2\right)^{1-\nu} \Theta(m-\lambda), \nl
\sigma(T) &=& 
    \frac{2}{\Gamma(\nu)} \left(\frac{\lambda}{2T}\right)^{\nu} K_{\nu}(\frac{\lambda}{T}). 
\ea{ANALYTIC-FAMILY}
with $0<\nu<2$. The $\nu=2$ limit belongs to a Dirac-delta mass distribution, the
$\nu=0$ limit to the Bessel function $K_0$ with logarithmic asymptotics for high
temperature (small argument). All these ansatze contain a mass-gap, the distributions
being zero for $m<\lambda$.
The $\nu=1/2$ value leads to the particularly simple eos: $\sigma=e^{-\lambda/T}$.

We found that an overall fit in the range of known lattice data
is also achieved by the analytic ansatz
\be
 \sigma(z) = \exp\left(- \lambda z  \right) \frac{1+e^{-a/b}}{1+e^{(z-a)/b}}
\ee{ANSATZ}
with $z=T_c/T$, $\lambda=1.05, a=0.90, b=0.11$ for data from~\cite{FODOR},
$\lambda=0.87, a=0.90, b=0.10$ for data from~\cite{BIELEFELD}
and $\lambda=0.56, a=0.83, b=0.10$ for data from~\cite{MILC}.
We note that in Refs.\cite{FODOR,BIELEFELD} $T_c\approx 170$ MeV, but in Ref.\cite{MILC} 
$T_c \approx 190$ MeV was taken.  
These fits are demonstrated in Fig.\ref{FIG-EOS}
where the different sets of lattice QCD data are compared with the 
fitted $ \sigma(z)=p/p_{SB} $ curves of eq.(\ref{ANSATZ}).

In this letter we investigate the lattice QCD eos data of Refs.~\cite{FODOR,MILC,BIELEFELD}
closely, but they qualitatively agree with other results on this issue.
The rise at moderately high temperatures (low $z$) 
cannot be accommodated by quantum statistical effects, 
but it can be  characterized as the effect of an exponential factor $exp(-\lambda T_c/T)$
in the range from $T_c$ to $2.5T_c$ (cf. Fig.\ref{FIG-EOS}).
While this moderately high-temperature behavior is well fitted by the pure
exponential $\sigma(z)$ function, the part below $T_c$ is more reduced.
The $\sigma(z)=1-K/ln(\eta/z)$ form is also able to fit $T > T_c$ data, but it goes to
negative values at a finite temperature, which is unphysical.
Our exponential fit is overall positive. 
A most satisfying extrapolation would interpolate between these two functions.


In Fig.\ref{FIG-EOS} lattice data from Ref.\cite{FODOR} (a), Ref.\cite{BIELEFELD} (b) and
Ref.\cite{MILC} (c) on $p/p_{SB} =  \sigma(z)$ as a function
of the temperature $T/T_c$ are plotted.
In a one-loop resummed pQCD motivated approach, using a mass directly proportional to the temperature
the approach to one is logarithmic, $1-K/\ln(\eta T/T_c)$ (dotted lines).
The exponential behavior, on the other hand, supports the presence
of a lowest mass in high-temperature QCD.
While this fact depends on the low-temperature drop of the pressure curve,
it is not easy to consolidate the effect due to quantitatively different
pressure curves presented by different lattice QCD calculations.
Although we do not intend to review lattice QCD eos calculations in this
paper, we note that the investigated simulations differ in the corresponding
value of the physical pion mass ($m_{\pi}\approx 140$ MeV for \cite{FODOR}, 
$m_{\pi}\approx 540$ MeV for \cite{BIELEFELD}
and $m_{\pi}\approx 300$ MeV for \cite{MILC}). There can be further differences of
technical nature, which we do not feel to be able to comment on.
In our further analysis we choose the data of the Budapest-Wuppertal group
\cite{FODOR} to seek for a corresponding mass distribution, but of course the same exercise
can be done for other sets of pressure data, too.

\begin{figure}
\begin{center}
 \includegraphics[width=0.33\textwidth,angle=-90]{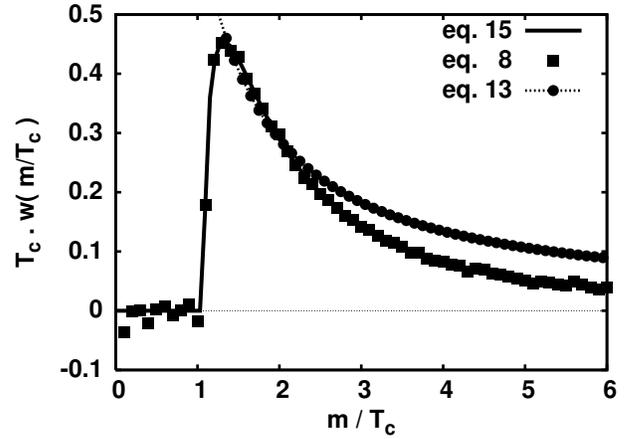}
 \end{center}
 \caption{ \label{FIG-fmt}  
  The mass distribution function obtained by evaluating the complex 
  integral from eq.(\ref{INV-MEIJER}) (boxes) and using the analytic fit 
  eq.(\ref{ANSATZ}) to lattice QCD eos data of Ref.\cite{FODOR}. The full
  line corresponds to eq.(\ref{ANALFT}), circles show the curve obtained
  by using the $1-K/ln(\eta T/T_c)$ type fit.
 }
 \end{figure}

To evaluate the integral given by eq.(\ref{INV-MEIJER}), we choose a simple
path parallel to the imaginary z axis, $ z = c + i \omega $. 
With numerical integration 
we obtain an $ f(t) $ mass distribution shown in Fig.\ref{FIG-fmt} by full boxes. 
Fluctuations at small masses are due to limitations of the applied numerical method. 
The part of the mass distribution shown here reconstructs the $T > T_c$
part of the pressure curve nicely, but it fails to approximate the pressure
at $T < T_c$. In the following we seek to understand this phenomenon. 

One can obtain simple analytic approximations for the $ f(t) $ function 
by expanding the expression for $ \sigma(z) $,  eq.(\ref{ANSATZ}).
However, requiring a convergent expansion, one arrives at two distinct 
series expansions: one  for $z<a$ and another one for $z>a$.
\ba
  \sigma(z) &=& e^{-z \, \lambda } + \ldots \, \, \, {\rm for} \, \, \,  z < a   \nl 
  \sigma(z) &=& e^{-z \, (\lambda + 1/b) } \,(1+e^{a/b}) + \ldots
   \, \, \, {\rm for} \, \, \,  z > a. 
\ea{SIGEXP}
The inverse Meijer K-transform of the simple exponential, $\exp(-\lambda \, z)$,
can be given based on an analytically known integral 
(cf. eq.\ref{ANALYTIC-FAMILY} for $\nu=1/2$):
\be
 f(t) = \frac{4\lambda}{t^2\pi}\sqrt{1-\frac{\lambda^2}{t^2}}.
\ee{ANALFT}
The above expression is valid for $t \ge \lambda$, for smaller $t=m/T_c$ values
$f(t)$ is identically zero. Hence the $t$-integration in the Meijer K-transform,
when determining the pressure contribution, starts at $t=\lambda$.
Physically this corresponds to a lowest mass in the continuous spectrum, to a
mass gap. Since both the approximations to $T \le T_c$ and to $T \ge T_c$ parts
of the pressure contain a leading exponential factor ($\lambda$ and
$\lambda + 1/b$ respectively, cf. eq.(\ref{SIGEXP})),
eos data seem to support a lowest value of a continuous mass spectrum both
in moderately low and moderately high temperature quark  matter 
(see the full line in Fig.\ref{FIG-fmt}).

Actually, requiring $z > a$ is equivalent to a Hagedorn limiting temperature
$T_H=T_c/a$, and in fact transforms back nearly 
to an exponentially rising mass spectrum part.  In this regime the QCD matter 
also has been fitted by a hadron resonance gas~\cite{HRESGAS}.

\begin{figure}[h]
\begin{center}
\includegraphics[width=0.33\textwidth,angle=-90]{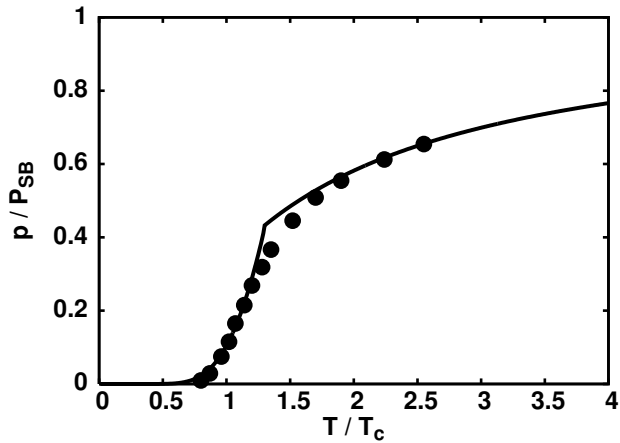}
\end{center}
\caption{ \label{FIG-press} 
  The normalized lattice QCD pressure and the pressure fitted to convergent series
  expansions, eq.(\ref{SIGEXP}), 
  obtained by numerically re-integrating $f(t)$ functions given by eq.(\ref{ANALFT}).
}
\end{figure}

Substituting the respective $f(t)$-s for $T<T_c$ and $T>T_c$ 
into eq.(\ref{ANALFT}) we calculate the  pressure from eq.(\ref{SIGMA}).
These two curves are shown in Fig.\ref{FIG-press}, together with the lattice QCD results
of Ref.\cite{FODOR}.
A numerical method designed to obtain an overall non-negative (probability like)
$f(t)$ distribution, which fits well some $\sigma(z_k)=s_k$ points,
is represented by the maximum entropy method (MEM). We applied this method to the
lattice QCD eos data discussed in this paper in order to obtain a mass distribution:
both by using a MEM program designed to invert the Meijer K-transform and also
by searching numerically for the inverse Laplace transform of $\sigma(z)$ first.
We failed, however, to obtain better numerical results then by evaluating the complex integral
eq.(\ref{INV-MEIJER}) as discussed above.

For calculating quark number susceptibilities or higher order Taylor coefficients
the Boltzmann approximation becomes unreliable; starting at the fourth order
the Boltzmannian term no longer dominates the Fermi distribution. Experience
with the hadronic resonance gas model supports the expectation that dependence on
chemical potential also can be interpreted in terms of clustered but non-interacting
components~\cite{HRESGAS,GORENSTEIN,BAGMASS}.

In conclusion, we have analyzed lattice QCD pressure data in terms of a continuous,
temperature independent mass distribution.
We find a strong indication for a finite mass gap in such quasiparticle models,
the details depending on the low temperature behavior of the pressure curve.
Since all simulation data are below the $M=2.5T_c$ curve, the immediate
conclusion would be that the $p(T)$ curve can be fitted by components with
higher mass only. Allowing for a milder drop of the pressure at low temperature
the lowest mass may be lower, we presented an example with $M\approx \lambda T_c$
with fitted $\lambda$-values near to one.
In general for any $p(T)$ curve showing finite $T^{n+1}$-weighted integrals for
$p/p_0$ the low-m behavior of $w(m)$ is restricted by finite integrals of $m^{-n}w(m)$.
Since a single-mass $p(T)/p_0$ curve cannot fit
the lattice QCD equation of state obtained by any of the groups calculating it, 
these data demand a finite width mass distribution.

For the physical problem of quark matter we have learned from the above analyses that either
the mass distribution is temperature dependent and then the thermodynamical description
is rather complex then, or there is a mass gap compatible to the equation of state unless
the pressure rises again at low temperatures (where we have presently no simulation data,
but the idea of a non-interacting pion gas would correspond to a pressure higher than zero).

{\bf Acknowledgment}

Enlightening discussions with Antal Jakov\'ac at BME 
about the maximum entropy method and related problems are hereby
gratefully acknowledged.
This work has been supported by the Hungarian National Science Fund OTKA
(T49466, T48489), by the EU-I3HP project and by a Bolyai scholarship
for P.~V\'an. 
%

\end{document}